\documentclass[10-pt]{article}

\begin{document}

\title{Extra force from an extra dimension. Comparison between brane theory, STM and other approaches}
\author{J. Ponce de Leon\thanks{E-mail: jponce@upracd.upr.clu.edu}\\ Laboratory of Theoretical Physics, Department of Physics\\ 
University of Puerto Rico, P.O. Box 23343, San Juan, \\ PR 00931, USA} 
\date{October  2003}

\maketitle
\begin{abstract}

 \end{abstract}
We investigate the question of how an observer in $4D$ perceives  the five-dimensional geodesic motion. We consider the interpretation of null and non-null bulk geodesics in the context of brane theory, space-time-matter theory (STM) and other non-compact approaches. We develop a ``frame-invariant" formalism that allows the computation of the rest mass and its variation as observed in $4D$. We find the appropriate expression for the four-acceleration and thus obtain the extra force observed in $4D$. 
Our formulae extend and generalize all previous results in the literature. 
An important result here is that the extra force in brane-world models with ${\bf Z}_{2}$-symmetry 
is continuous and well defined across the brane. 
This is because the momentum component along the extra dimension is discontinuous across the brane, 
which effectively compensates the discontinuity of the extrinsic curvature. We show that brane theory and STM produce identical interpretation of the bulk geodesic motion. This holds for null and non-null bulk geodesics. Thus, experiments with test particles are unable  to distinguish  whether our universe  is described by the brane world scenario or by STM.  However, they do discriminate between the brane/STM scenario and other non-compact approaches. Among them the canonical and embedding approaches, which we examine in detail here.



\newpage
\section{Introduction}

The concept  that our world may be embedded in a universe that possesses more than four dimensions has a long and distinguished history. In theoretical physics, it can be traced back to the pioneers works of Kaluza \cite{Kaluza} and Klein \cite{Klein} who  interpreted the electromagnetic field as a geometrical effect of a hidden fifth dimension. Currently, theories of the Kaluza-Klein type in many dimensions are used in different branches of physics. Superstrings ($10D$) and supergravity $(11D)$ are well known examples \cite{collins}.

 In higher-dimensional gravity theories, the scenario is that matter fields are confined to our four-dimensional universe, a $3$-brane, in a $1 + 3 + d$ dimensional spacetime, while gravity propagates in the extra $d$ dimensions as well \cite{Arkani1}-\cite{Arkani3}. In these theories there are several motivations for the introduction of extra dimensions. Among them to resolve the differences between  gravity and quantum field theory and ultimately unify all forces of nature. Also, as providing possible solutions to the hierarchy and the cosmological constant problems \cite{RS1}-\cite{RS2}.

The idea of extra dimensions is also inspired by the vision that  matter in $4D$ is purely geometric in nature. In space-time-matter theory (STM) one {\em large} extra dimension is needed in order to get a consistent description, at the macroscopic level, of the properties of  the matter as observed in $4D$ \cite{JPdeL 1}-\cite{EMT}. The mathematical support of this theory is given by a theorem of differential geometry due to Campbell and Magaard \cite{Campbell}-\cite{Lidsey}. 

Although these theories have different motivations for the introduction of extra dimensions,  
they confront similar challenges. From a theoretical viewpoint, they have to  predict observable effects from the extra dimensions. From an experimental viewpoint, the vital issue is the discovery of new physical phenomena, which could unambiguously be associated with the existence of extra dimensions. 

A possible way of testing for new physics coming from  extra dimensions is to  examine the dynamics of test particles. In practice this means to search for deviations from the universal  ``free fall" in $4D$. For that reason the geodesic motion on $5D$ manifolds and $4D$ submanifolds has been a subject of intensive investigations \cite{MashhoonWesson}-\cite{Romero2}. Two main results have emerged  from the dimensional reduction of geodesics in $5D$. Firstly,  that the free motion in $5D$ is observed in $4D$ as being under the influence of a non-gravitational force, if the velocity of the test particle has non-zero component along the extra dimension.  Secondly, since the extra force has a component which is parallel to the particle's four-velocity, the rest mass is observed to vary with time. 

These results  are important in view of their potential experimental/observational relevance. However, their interpretation  and the new physics related to them,  is not clear  yet. In fact, despite some successful applications,  the implementation of these results in the context of brane theory and STM has lead to a number of  statements and conclusions that we believe should be reconsidered.  

One of them  is that the extra force {\em cannot} be implemented directly in brane-world models, in the RS2 scenario \cite{RS2}, because the derivatives of the metric are discontinuous, and change sign, through the brane (see for example \cite{Youm2}-\cite{Seahra3}). A related statement is that the extra force is zero in brane-world cosmological models with ${\bf Z}_{2}$ symmetry. 

In this work we bring  a positive perspective to this topic. We demonstrate that, in brane-world models with ${\bf Z}_{2}$-symmetry,
the extra force is continuous and well defined across the brane. 
We show, by means of explicit examples, that the extra force in cosmological models with ${\bf Z}_{2}$ symmetry is {\em not} necessarily zero.  We also illustrate how the force and mass,  as observed on the three-brane, depend on whether the  bulk motion is along null or  non-null geodesics. 

Another  new discovery in this work is that brane theory and STM produce the {\em same} results for test particles as observed in $4D$. Consequently, for the computation of the extra force and mass we can ignore the details of whether the bulk geodesic motion is interpreted on the non-singular hypersurface of STM or on the singular hypersurface of brane theory.  This equivalence has nothing to do with the dynamics in $4D$, but it is a result of the assumption that test particles move along five-dimensional geodesics in both theories. From an observational  viewpoint, this means that experiments measuring the extra force acting on test particles are not able  to discriminate whether our universe  is described by the brane world scenario or by STM. In order to settle this point,  a self-consistent analysis of the combination of physical, astrophysical and cosmological effects like in Ref. \cite{GLH}  should be made.  

 We also elucidate some important issues related to the interpretation of STM in $4D$. We refer to  the concept  that the extra force can be made to dissappear by changing the parameterization of the metric.  This seems to be related to the geodesic approach where the mass of the particle depends on the affine parameters used to describe the motion in $5D$ and $4D$ \cite{Seahra4}.  

In this work we show how the rest mass as well as the  extra force as observed in $4D$ crucially depend on the method  we use to identify  the $4D$ metric from the $5D$ one. 
In particular we point out that, unlike the case of brane theory, each bulk metric in STM can be used to generate at least {\em six} different expressions  for the mass and force as observed in $4D$ (for null and non-null bulk geodesics). This wealth of interpretations is not a consequence of changing any parameter in the bulk metric, but it is an attribute  of STM, where the number of physical restrictions in the theory is not in general sufficient to determine the properties in $4D$ \cite{STMBrane}. 

We also clarify the question of whether the extra force is a pure consequence of the fact that the bulk metrics in brane theory and STM are allowed to depend on the extra coordinate. This  is certainly true when the metric along the fifth dimension is flat.  However,  in general we find that a large extra dimension {\em does not} necessarily imply the existence of an extra non-gravitational force. Conversely, in general a compact extra dimension does not preclude the existence of an extra force. 

The structure of the paper is as follows: In section $2$ we recall the definition of the relativistic force four-vector in covariant and contravariant components. 
We also recall some requirements  on the covariant derivative in $4D$. In section $3$ we present the bulk metric and develop a ``frame-invariant" formalism that allows the computation of rest mass and its variation as observed in $4D$. In section $4$ we expound some technical problems which arise when  the covariant derivative  in $5D$  is used in $4D$. Next, we  define  the appropriate covariant derivative in $4D$ and find the four-acceleration, which satisfies physical conditions. This allows us to find the four-force in agreement with the definition given in section $2$. Throughout the discussion we consider the interpretation of null and non-null geodesic motion in the bulk. In sections $5$ and $6$ we apply our formalism to the brane-world scenario and STM respectively.  In section $7$ we discuss the canonical metric and the foliating approach as alternative interpretations to STM. Finally, in section $8$ we give a summary.

\section{Definition of force in $4D$}
Here we present the definition of relativistic four-force that we are going to use throughout this paper. 
In four dimensions the motion of a test particle is described by its four velocity 
\begin{equation}
u^{\mu} = \frac{dx^{\mu}}{ds}, \;\;\;\;\; u_{\mu}u^{\mu} = 1.
\end{equation}
The four-momentum of a particle of rest mass $m_{0}$ is defined as 
\begin {equation}
p^{\mu} = m_{0}u^{\mu}, \;\;\;\;\;p_{\mu} = m_{0}u_{\mu}.
\end{equation}
In special relativity, in Cartesian coordinates the four-force acting on a test particle is given by
\begin{equation}
\label{force in Cartesian coordinates 1}
F^{\mu} = \frac{d p^{\mu}}{ds} = \frac{d}{ds}(m_{0}u^{\mu}).
\end{equation}
Thus, 
\begin{equation}
\label{force in special relativity}
\label{force in Cartesian coordinates 2}
\frac{F^{\mu}}{m_{0}} = \frac{du^{\mu}}{ds} + \frac{u^{\mu}}{m_{0}}\frac{dm_{0}}{ds}.
\end{equation}
If the rest mass of the particle is constant along its motion, then the $4D$-force is orthogonal to the four velocity, i.e.,  $F^{\mu}u_{\mu} = 0$. Otherwise, the four-force has a component parallel to the four-velocity such that $F^{\mu}u_{\mu} = dm_{0}/ds$. 

In curvilinear coordinates the metric of the spacetime is described by a symmetric tensor $g_{\mu\nu}$. In such coordinates, the appropriate  generalization of (\ref{force in special relativity}) is 
\begin{equation}
\label{contravariant force in curvilinear coordinates}
\frac{F^{\mu}}{m_{0}} = \frac{D^{(4)}u^{\mu}}{ds} + \frac{u^{\mu}}{m_{0}}\frac{dm_{0}}{ds},
\end{equation}
where $D^{(4)}$ denotes the covariant differential calculated in $4D$, i.e., 
\begin{equation}
\label{prop 1 of D4}
D^{(4)}g_{\mu\nu} = 0.
\end{equation}
The indexes of four-vectors and four-tensors are lowered and raised with the aid of  $g_{\mu\nu}$. For instance,
\begin{equation}
u_{\mu} = g_{\mu\nu}u^{\nu}, \;\;\;\; F_{\mu} = g_{\mu\nu} F^{\mu}. 
\end{equation}
Consequently, 
\begin{equation}
\label{prop 2 of D4}
u_{\mu}D^{(4)}u^{\mu} = u^{\mu}D^{(4)}u_{\mu} = 0,
\end{equation}
and the covariant components of the four-force are given by 
\begin{equation}
\label{covariant force in curvilinear coordinates}
\frac{F_{\mu}}{m_{0}} = \frac{D^{(4)}u_{\mu}}{ds} + \frac{u_{\mu}}{m_{0}}\frac{dm_{0}}{ds},
\end{equation}
In this way 
\begin{equation}
\label{condition on force}
F^{\mu}u_{\mu} = F_{\mu}u^{\mu} = \frac{dm_{0}}{ds},
\end{equation}
is valid not only in Cartesian coordinates, but in {\em all} coordinate systems. We will use these properties in section $4$.
\section{Motion in higher dimensions}

In the Randall-Sundrum brane-world scenario and other non-compact Kaluza-Klein theories,  the motion of test particles is higher-dimensional in nature. In other words, all test particles travel on five-dimensional geodesics but observers, who are bounded to spacetime, have access only to the $4D$ part of the trajectory. From a mathematical viewpoint, this means that the equations governing the motion in $4D$ are projections of the $5D$ equations on the $4D$-hypersurfaces orthogonal to some vector field $\psi^A$. The corresponding projector can be written as
\begin{equation}
\label{projector}
h_{AB} = \gamma_{AB} - \epsilon \psi_{A}\psi_{B}, 
\end{equation}
where $\gamma_{AB}$ is the five-dimensional metric and the factor $\epsilon$ can be $- 1$ or $+ 1$ depending on whether the extra dimension is spacelike or timelike, respectively. 
In what follows we will consider the background $5D$ metric 
\begin{equation}
d{\cal{S}}^2 = {\gamma}_{\mu\nu}(x^{\rho}, y)dx^{\mu}dx^{\nu} + \epsilon \Phi^2(x^{\rho}, y)dy^2,
\end{equation}
where $\gamma_{\mu\nu}$ is the metric {\em induced} in $4D$.  The vector $\psi^A$, orthogonal to spacetime is given by
\begin{equation}
\label{e4}
{\psi^A}= (0, 0, 0, 0, {\Phi}^{- 1}),\;\;\;\;\;\psi_{A}\psi^{A} = \epsilon.
\end{equation}
In order to obtain the four-dimensional interpretation of the geodesic motion in $5D$, we have to decide how to identify the physical or observable spacetime metric from the induced one. In brane-world  theory and STM the spacetime metric  $g_{\mu\nu}$ is commonly identified with  $\gamma_{\mu\nu}$. However, in some  approaches the physical metric in $4D$ is assumed to be conformally related to the induced one, viz.,
\begin{eqnarray}
\label{general metric}
d{\cal{S}}^2 &=& \Omega(y)g_{\mu\nu}(x^{\rho}, y)dx^{\mu}dx^{\nu} + \epsilon \Phi^2(x^{\rho}, y)dy^2,\nonumber \\
&=& \Omega(y) ds^2 + \epsilon \Phi^2(x^{\rho}, y) dy^2,   
\end{eqnarray}
where $\Omega(y)$ is called ``warp" factor and satisfies the obvious condition that  $\Omega > 0$. This line element is  more general than  the Randall-Sundrum metric, the so-called canonical metric,  and encompasses all the metrics generally used in brane-world and STM theories. The object of this section is to examine the motion of test particles in the background metric (\ref{general metric}). 

In order to facilitate the discussion and make the presentation self-consistent, we will give a brief review of our formalism  \cite{DynKK} for the effective rest mass $m_{0}$, and its variation along the observed trajectory in $4D$,  as an effect caused by the motion (momentum) along the extra dimension.  Some  technical details of the discussion depend on whether the test particle in $5D$ is massive or massless. We  therefore approach these two cases separately. 

\subsection{Massive particles in $5D$} 

Let us consider a massive test particle moving in a five-dimensional manifold with metric (\ref{general metric}).
The momentum $P^A$ of such a particle (extending the dynamics of test particles from $4D$ to $5D$) is defined in the usual way, namely,
\begin{equation}
\label{5D Momentum}
P^{A} = M_{(5)}\left(\frac{dx^{\mu}}{d{\cal{S}}}, \frac{dy}{d{\cal{S}}}\right), 
\end{equation}
where $M_{(5)} > 0$ is the constant five-dimensional mass of the particle and $U^A = (dx^{\mu}/d{\cal{S}}, dy/d{\cal{S}})$ is the velocity in $5D$. Thus $U^AU_A = 1$ and 
\begin{equation}
\label{5D mass}
P^{A}P_{A} = M_{(5)}^2. 
\end{equation}
We note that five-dimensional indexes  are lowered and raised with the aid of the $5D$ metric $\gamma_{AB}$. 

The five-dimensional motion is perceived by an observer in $4D$ as the motion of a particle with four-momentum $p_{\mu}$. Consequently,  the  effective rest mass in $4D$ is given by
\begin{equation}
\label{4D mass}
p_{\alpha}p^{\alpha} = m_{0}^2,
\end{equation}
where the four-dimensional indexes are lowered and raised by the spacetime metric $g_{\mu\nu}$.
Because of the absence of cross terms in (\ref{general metric}), the $4D$ components  of $P_{A}$ and $P^{A}$ (i.e., $A = 0, 1,2,3$) are already ``projected" onto spacetime. Namely,
\begin{equation}
\label{identification of p with P cov}
p_{\mu} = h_{\mu A}P^{A} = h_{\mu\nu}P^{\nu} = \Omega g_{\mu\nu}P^{\nu} = P_{\mu}.
\end{equation}
Thus from (\ref{5D mass}) we get  
\begin{equation}
\label{rel between m, P4 and M}
m_{0}^2 + \Omega(y)P_{4}P^{4} = \Omega(y)M_{(5)}^2.
\end{equation}
Therefore,  the relation between the rest mass in $4D$ and $5D$ is given by
\begin{equation}
\label{relation between the rest mass in 4D and 5D}
m_{0} = \sqrt{\Omega}M_{(5)}\left[1 + \frac{\epsilon \Phi^2}{\Omega}\left(\frac{dy}{ds}\right)^2\right]^{- 1/2}.
\end{equation}
This equation is the five-dimensional counterpart to  $m = m_{0}[1 - v^2]^{-1/2}$, for the variation of particle's mass due to its motion in spacetime. It shows how the nature of the extra dimension and the motion   in $5D$  affect the rest mass measured in $4D$.  It allows us to conclude that  $m_{0}$ depends on (i) the mass of the particle in $5D$, (ii) the character of motion in $5D$, i.e. on $dy/ds$, and (iii) the nature of the extra coordinate, i.e., whether it is spacelike or timelike. 

\subsubsection{Variation of rest mass for $M_{(5)} \neq 0$}

From (\ref{rel between m, P4 and M}) it follows that if the trajectory in $5D$ lies entirely on a hypersurface $y = const$, i.e. if $P^4 = P_4 = 0$, then the observed mass in $4D$ is constant. The opposite happens if the five-dimensional motion has non vanishing velocity along $y$. In this case the rest masses,  measured  by an observer in $4D$, in general vary along the trajectory.

In order to find the observed variation of $m_{0}$ we have to evaluate $dm_{0}/ds$. This requires the computation  of $dP^4/ds$ and $dP_4/ds$,  which can be easily done from  the geodesic equation in $5D$,
\begin{equation}
\label{5D geodesic}
\frac{dU^{A}}{d{\cal{S}}} + K^{A}_{BC}U^{B}U^{C} = 0,
\end{equation}
where $U^A = (dx^{\mu}/d{\cal{S}}, dy/d{\cal{S}})$ is the five-velocity and $K^{A}_{BC}$ is the Christoffel symbol formed with the $5D$ metric $\gamma_{AB}$.  We also have to use the relationship 
\begin{equation}
\label{rel between  dS and ds}
\frac{d{\cal{S}}}{M_{(5)}} = \Omega \frac{ds}{m_{0}},
\end{equation}
which follows from (\ref{general metric}) and  (\ref{relation between the rest mass in 4D and 5D}). 

Thus, setting $A = 4$ in (\ref{5D geodesic}) we obtain  
\begin{equation}
\label{variation of P4 contr}
\frac{1}{m_{0}}\frac{dP^{4}}{ds} = \frac{\epsilon}{2\Omega \Phi^2}\frac{\partial(\Omega g_{\mu \nu})}{\partial y} u^{\mu}u^{\nu} - \frac{2 u^{\mu}}{\Omega \Phi}\frac{\partial \Phi}{\partial x^{\mu}}\left(\frac{dy}{ds}\right) - \frac{1}{\Omega \Phi}\frac{\partial \Phi}{\partial y}\left(\frac{dy}{ds}\right)^2,
\end{equation}
where $u^{\mu} = (dx^{\mu}/ds)$ is the usual four-velocity of the particle. Also, for the covariant component we get
\begin{equation}
\label{variation of P4}
\frac{1}{m_{0}}\frac{dP_{4}}{ds} = \frac{1}{2\Omega}\frac{\partial(\Omega g_{\mu \nu})}{\partial y} u^{\mu}u^{\nu} +  \frac{\epsilon \Phi}{\Omega}\frac{\partial \Phi}{\partial y}\left(\frac{dy}{ds}\right)^2,
\end{equation}

Now, taking derivative of (\ref{rel between m, P4 and M}) and using the above expressions 
(\ref{variation of P4 contr}), (\ref{variation of P4}) we obtain the variation of the effective rest mass as follows
\begin{equation}
\label{variation of the effective mass}
\frac{1}{m_{0}}\frac{dm_{0}}{ds} = - \frac{1}{2}u^{\mu}u^{\nu}\frac{\partial g_{\mu\nu}}{\partial y} \left(\frac{dy}{ds}\right) + \frac{\epsilon \Phi u^{\mu}}{\Omega}\frac{\partial \Phi}{\partial x^{\mu}}\left(\frac{dy}{ds}\right)^2.
\end{equation}
\subsection{Massless particles in $5D$}

Let us now consider massless test particles, $M_{(5)} = 0$,  moving in the five-dimensional metric (\ref{general metric}). 
The motion of such particles is along isotropic geodesics, which in five-dimensions  requires $d{\cal{S}} = 0$.  Therefore, \begin{equation}
\label{ds for null geodesics}
\Omega ds^2 = - \epsilon \Phi^2 dy^2.
\end{equation}
It is  clear that the signature of the extra dimension plays an important role here. In particular, null geodesics in $5D$ appear as timelike paths in $4D$ only if the following two conditions are met {\em simultaneously}: (i) the extra dimension is spacelike, and (ii) the particle in its five-dimensional motion has $P_{4} \neq 0$. Otherwise, a null geodesic in $5D$ is observed as a lightlike particle in $4D$.

 In the case where $M_{(5)} = 0$, the derivatives $M_{(5)}d/d{\cal{S}}$ in (\ref{5D Momentum})   have to be replaced by $d/d\lambda$, where $\lambda$ is the parameter along the null $5D$ geodesic \cite{Landau and Lifshitz}. Thus, from (\ref{rel between m, P4 and M}), with $M_{(5)} = 0$, $\epsilon = - 1$ and  $P^{4} = dy/d\lambda$,  we obtain
\begin{equation}
\label{m and P4 for massless particles in 5D}
m_{0} = \pm \sqrt{\Omega} \Phi \frac{dy}{d \lambda} = \mp \frac{\sqrt{\Omega}}{\Phi}P_{4} > 0.
\end{equation}
It is important to mention that $P_{4}$ is independent of $\lambda$, which means that the mass calculated from  (\ref{m and P4 for massless particles in 5D}) is unaffected by the parameterization along the five-dimensional null geodesic.  This can be clearly illustrated in terms of the five-dimensional action $S$, in which case  $m_{0} = (\sqrt{\Omega}/\Phi) |\partial S/ \partial y|$. We will come back to this point in sections $5$ and $7$.
\subsubsection{Variation of rest mass for $M_{(5)} = 0$}
 
From (\ref{ds for null geodesics}), for a spacelike extra coordinate $(\epsilon = -1)$,  it follows that\footnote{When taking the roots we choose the signs in such a way that $m_{0} > 0$ and  $d\lambda/ds >0$.} $dy =  \pm (\sqrt{\Omega}/\Phi)ds$. Therefore, 
\begin{equation}
\label{geodesic parameter and rest mass}
d \lambda = \left(\frac{\Omega}{m_{0}}\right)ds.
\end{equation}
From this and  the $4$-component of the geodesic equation we obtain
\begin{equation}
\frac{1}{m_{0}}\frac{dP_{4}}{ds} = \frac{1}{2\Omega}\frac{\partial(\Omega g_{\mu \nu})}{\partial y} u^{\mu}u^{\nu} - \frac{1}{\Phi}\frac{\partial \Phi}{\partial y}.
\end{equation}
Consequently, the variation of rest mass for a spacelike extra coordinate $(\epsilon = -1)$ and $M_{(5)} = 0$  is obtained from (\ref{m and P4 for massless particles in 5D}),  as 
\begin{equation}
\label{variation of effective rest mass for massless 5D particles}
\frac{1}{m_{0}}\frac{dm_{0}}{ds} = \mp \frac{\sqrt{\Omega}}{2 \Phi}\frac{\partial g_{\mu\nu}}{\partial y}u^{\mu}u^{\nu} - \frac{u^{\mu}}{\Phi}\frac{\partial \Phi}{\partial x^{\mu}}. 
\end{equation}
We note that although the mathematical description of massless particles in $5D$ differs from that of massive particles in $5D$, the last two equations (for $M_{(5)} = 0$) can be readily obtained from (\ref{variation of P4}) and (\ref{variation of the effective mass}) (for $M_{(5)} \neq 0$) just by setting $(dy/ds) = \pm (\sqrt{\Omega}/\Phi)$ and $\epsilon = -1$. 

We emphasize that for a timelike extra dimension $(\epsilon = +1)$, there is only one physical possibility. Namely, massless particles in $5D$  are perceived as massless particles in $4D$.  In addition,  their motion  is confined to hypersurfaces $y = const$. 

\medskip

To summarize, a bulk test particle moving freely in a five-dimensional manifold is observed in $4D$ as a test particle with variable rest, as given by (\ref{variation of the effective mass}) or (\ref{variation of effective rest mass for massless 5D particles}). We would like to emphasize that this is {\em not}  an artifact of a poor choice of coordinates or parameter used in the geodesic description, but it is a genuine four-dimensional manifestation of the extra dimension.

\section{Dynamics of test particles from $5D$ to $4D$}

So far we have only used the fourth component of the five-dimensional geodesic equation (\ref{5D geodesic}). We now turn our attention to the spacetime components of that  equation. 

Setting  $A = \mu$ in (\ref{5D geodesic}) and using  
\begin{equation}
U^{\mu} = \frac{m_{0}}{\Omega M_{(5)}} u^{\mu}, \;\;\;\;\;U_{\mu} = \frac{m_{0}}{M_{(5)}}u_{\mu}, 
\end{equation}
we find
\begin{equation}
\label{f lit contr}
\frac{Du^{\mu}}{ds} \equiv \frac{d u^{\mu}}{ds} + \Gamma_{\alpha \beta}^{\mu}u^{\alpha}u^{\beta} = \left(\frac{1}{2}u^{\mu}u^{\rho} - g^{\mu\rho}\right)u^{\lambda}\frac{\partial{g_{\rho\lambda}}}{\partial{y}}\left(\frac{dy}{ds}\right) + \frac{\epsilon \Phi}{\Omega}\left[\Phi^{;\mu} - u^{\mu}u^{\rho}\Phi_{;\rho}\right]\left(\frac{dy}{ds}\right)^2,
\end{equation}
where $\Gamma_{\alpha \beta}^{\mu}$ is the Christoffel symbol calculated with  the spacetime metric $g_{\alpha\beta}$.

For the covariant components of the four-velocity we get,
\begin{equation}
\label{f lit cov}
\frac{Du_{\mu}}{ds} \equiv \frac{d u_{\mu}}{ds} -  \Gamma_{\mu \alpha}^{\beta}u^{\alpha}u_{\beta}  = \frac{1}{2}u_{\mu}u^{\lambda}u^{\rho}\frac{\partial {g_{\lambda\rho}}}{\partial y}\left(\frac{dy}{ds}\right)  + \frac{\epsilon \Phi}{\Omega} \left[\Phi_{;\mu} - u_{\mu}\Phi_{;\rho}u^{\rho} \right](\frac{dy}{ds})^2.
\end{equation}
Thus, for any given five-dimensional metric (\ref{general metric}), we can always (at least in principle) solve the above equations  to find 
\begin{equation} 
u^{\mu} = u^{\mu}(s), 
\end{equation}
and the observed trajectory in $4D$
\begin{equation}
\label{4D trajectory}
x^{\mu} = x^{\mu}(s).
\end{equation}

From (\ref{f lit contr}) and (\ref{f lit cov}) it is clear that the geodesic motion in the five-dimensional manifold is observed in $4D$ to be under the influence of  extra non-gravitational forces.

\medskip

At first glance one would identify these extra forces with the terms on the right hand side of (\ref{f lit contr}) and (\ref{f lit cov}), and this is indeed the usual approach. However, this identification faces two problems. 

\medskip

First of all, while the observed force should be a four-vector, the quantities $Du^{\mu}/ds$ and $Du_{\mu}/ds$ given above do  {\em not} represent the contravariant and covariant component of any four-dimensional vector. In order to see this, let us notice that 
\begin{equation}
\label{bad vectorial properties of f lit 1}
u_{\mu}\left(\frac{Du^{\mu}}{ds}\right) \neq u^{\mu}\left(\frac{Du_{\mu}}{ds}\right) \neq 0. 
\end{equation}
Clearly this is not what we expect in $4D$, which is given by (\ref{prop 2 of D4}) as a result of $u_{\mu}u^{\mu} = 0$. 

The second delicate point  here is that
\begin{equation}
\label{bad vectorial properties of f lit 2}
g_{\mu\nu}\left(\frac{Du^{\mu}}{ds}\right) = \left(\frac{Du_{\nu}}{ds}\right) - u^{\lambda}\frac{\partial{g_{\nu\lambda}}}{\partial{y}}\left(\frac{dy}{ds}\right). 
\end{equation}
Thus, condition (\ref{condition on force}) is not satisfied if we identify  the r.h.s. of (\ref{f lit contr}) and (\ref{f lit cov}) with the contravariant and covariant components of the extra forces.  On the other hand, if $Du^{\mu}/{ds}$ and $Du_{\mu}/{ds}$ were the contravariant and covariant components of a four-vector\footnote{We note that in the case of Kaluza-Klein theories with the so-called ``cylinder" condition, (i.e. $\partial{g_{\mu\nu}}/\partial{y} = 0$) the quantities  $(Du^{\mu}/{ds})$ and $Du_{\mu}/{ds}$ do represent  the contravariant and covariant components of a four-vector, namely the four-acceleration}, then  they would comply with $g_{\mu\nu}(Du^{\mu}/{ds}) = (Du_{\nu}/{ds})$.  

Third, the appropriate definition of force should involve the change of particle's momentum, as in (\ref{force in Cartesian coordinates 1}) and (\ref{force in Cartesian coordinates 2}).

\subsection{The four-acceleration} 

Thus, the direct identification of $(Du^{\mu}/{ds})$ and $(Du_{\mu}/{ds})$ with the contravariant and covariant components of the extra non-gravitational force is questionable. Meanwhile, the correct definition of force (per unit mass) in $4D$, free of the problems mentioned above, was discussed in section $2$. 
It contains two terms; 
one of them is $(u^{\mu}/m_{0})(dm_{0}/ds)$, which has  already been obtained in (\ref{variation of the effective mass}) and/or  (\ref{variation of effective rest mass for massless 5D particles}). 
The other term has yet to be found; it is the four-acceleration $D^{(4)}u^{\mu}/ds$, which is the same for all test particles regardless of their mass.

The vectorial nature of the force in (\ref{contravariant force in curvilinear coordinates}) and (\ref{covariant force in curvilinear coordinates}), is assured by the fact that $D^{(4)}g_{\mu\nu} = 0$. On the other hand, 
\begin{equation}
\label{D of 4D metric}
Dg_{\mu\nu} = \frac{\partial{g_{\mu\nu}}}{\partial{y}}dy.
\end{equation}
This means that the operator $D$ defined in (\ref{f lit contr}) and (\ref{f lit cov}) is {\em not} the appropriate covariant differential calculated in $4D$. 
In order to construct the appropriate differential in $4D$, let us notice that
\begin{equation}
\frac{\partial{u^{\mu}}}{\partial y} = - \frac{1}{2}u^{\mu}u^{\alpha}u^{\beta}\frac{\partial{g_{\alpha\beta}}}{\partial y}, \;\;\;\; u^{\mu}\frac{\partial{u_{\mu}}}{\partial y} = \frac{1}{2}u^{\alpha}u^{\beta}\frac{\partial{g_{\alpha\beta}}}{\partial y},
\end{equation}
which can be easily shown in the comoving frame of reference. Using these expressions, from (\ref{f lit contr}), (\ref{f lit cov}) and (\ref{D of 4D metric}) we get
\begin{equation}
u_{\mu}\left[\frac{D}{ds} - \frac{dy}{ds}\frac{\partial}{\partial y}\right]u^{\mu} = 0, \;\;\;u^{\mu}\left[\frac{D}{ds} - \frac{dy}{ds}\frac{\partial}{\partial y}\right]u_{\mu} = 0, \;\;\;\left[\frac{D}{ds} - \frac{dy}{ds}\frac{\partial}{\partial y}\right]g_{\mu\nu} = 0.
\end{equation}
If we compare these expressions with (\ref{prop 1 of D4}) and (\ref{prop 2 of D4}) it is clear that a suitable definition for  $D^{(4)}$ is given by
\begin{equation}
\frac{D^{(4)}}{ds} \equiv \left[\frac{D}{ds} - \frac{dy}{ds}\frac{\partial}{\partial y}\right]. 
\end{equation}
For the case of more general metrics, $D^{(4)}$ can also be defined, but this requires the introduction of the appropriate projectors \cite{Ponce de Leon 1}.

With this definition we have 
\begin{equation}
u_{\mu}\frac{D^{(4)}u^{\mu}}{ds} = 0, \;\;\;u^{\mu}\frac{D^{(4)}u_{\mu}}{ds} = 0, \;\;\;\frac{D^{(4)}g_{\mu\nu}}{ds} = 0.
\end{equation}
As a consequence the acceleration is a four-vector. Namely, from (\ref{f lit contr}) we get 
\begin{equation}
\label{D4 of u contr}
\frac{D^{(4)}u^{\mu}}{ds} = \left(u^{\mu}u^{\rho} - g^{\mu\rho}\right)u^{\lambda}\frac{\partial{g_{\rho\lambda}}}{\partial{y}}\left(\frac{dy}{ds}\right) + \frac{\epsilon \Phi}{\Omega}\left[\Phi^{;\mu} - u^{\mu}u^{\rho}\Phi_{;\rho}\right]\left(\frac{dy}{ds}\right)^2.
\end{equation}
On the other hand, from (\ref{f lit cov}) we obtain
\begin{equation}
\label{D4 of u cov}
\frac{D^{(4)}u_{\mu}}{ds} = \left(u_{\mu}u^{\rho} - \delta_{\mu}^{\rho}\right)u^{\lambda}\frac{\partial{g_{\rho\lambda}}}{\partial{y}}\left(\frac{dy}{ds}\right) + \frac{\epsilon \Phi}{\Omega} \left[\Phi_{;\mu} - u_{\mu}\Phi_{;\rho}u^{\rho} \right](\frac{dy}{ds})^2.
\end{equation}
Clearly $g_{\mu\nu}(D^{(4)}u^{\mu}/ds) = (D^{(4)}u_{\nu}/ds)$ as required for the correct vectorial behavior of the four-acceleration.
\newpage
\subsection{The extra force observed in $4D$}
Collecting results we obtain the explicit form of the extra force as follows.

\paragraph{Massive particles in $5D$:} Using the definitions (\ref{contravariant force in curvilinear coordinates}) and (\ref{covariant force in curvilinear coordinates}), we find that a massive bulk test particle ($M_{(5)} \neq 0$) moving freely in a five-dimensional manifold is observed in $4D$ as a massive particle ($m_{0} \neq 0$)  moving under the influence of the force
\begin{equation}
\label{new def for force per unit mass contr}
\frac{1}{m_{0}}F^{\mu} = \frac{D^{(4)}u^{\mu}}{ds} + \frac{u^{\mu}}{m_{0}}\frac{dm_{0}}{ds} = \frac{\epsilon \Phi}{\Omega} \Phi^{\mu}\left(\frac{dy}{ds}\right)^2 + \left(\frac{1}{2}u^{\mu}u^{\rho} - g^{\mu\rho}\right) u^{\lambda}\frac{\partial g_{\rho \lambda}}{\partial y} \frac{dy}{ds}.
\end{equation}
The covariant components are
\begin{equation}
\label{new def for force per unit mass cov}
\frac{1}{m_{0}}F_{\mu} = \frac{D^{(4)}u_{\mu}}{ds} + \frac{u_{\mu}}{m_{0}}\frac{dm_{0}}{ds} = \frac{\epsilon \Phi}{\Omega} \Phi_{\mu}\left(\frac{dy}{ds}\right)^2 + \left(\frac{1}{2}u_{\mu}u^{\rho} - \delta_{\mu}^{\rho}\right) u^{\lambda}\frac{\partial g_{\rho \lambda}}{\partial y} \frac{dy}{ds}.
\end{equation}
We see that the extra force is made up of three distinct contributions, viz.,  

\begin{equation}
F^{\mu} = F^{\mu}_{\Phi \perp} + F^{\mu}_{g \perp} + F^{\mu}_{\parallel},
\end{equation}
where
\begin{equation}
\frac{1}{m_{0}}F^{\mu}_{\Phi \perp} = \frac{\epsilon \Phi}{\Omega} [\Phi^{\mu} - u^{\mu}\Phi_{\rho}u^{\rho}]\left(\frac{dy}{ds}\right)^2,
\end{equation}
\begin{equation} 
\frac{1}{m_{0}}F^{\mu}_{g \perp} = [u^{\mu}u^{\rho} - g^{\mu\rho}] u^{\lambda}\frac{\partial g_{\rho \lambda}}{\partial y} \frac{dy}{ds},
\end{equation}
and 
\begin{equation}
\label{extra force parallel to four velocity, new definition}
\frac{1}{m_{0}}F^{\mu}_{\parallel} = u^{\mu}\left[- \frac{1}{2}u^{\lambda}u^{\rho}\frac{\partial g_{\lambda\rho}}{\partial y} \frac{dy}{ds} + \frac{{\epsilon \Phi u^{\lambda}}}{\Omega}\frac{\partial \Phi}{\partial x^{\lambda}}\left(\frac{dy}{ds}\right)^2\right].
\end{equation}
All these terms have to be evaluated along the trajectory (\ref{4D trajectory}). We note that $F^{\mu}_{\Phi \perp}$ and  $F^{\mu}_{g \perp}$ are orthogonal to the four-velocity, while $F^{\mu}_{\parallel}$ is parallel to it. They crucially depend on the motion along the extra dimension. In particular,  if the $5D$ motion is confined to hypersurfaces with $y = const$, then $F^{\mu}_{\Phi \perp} = F^{\mu}_{g \perp} = F^{\mu}_{\parallel} = 0$, identically.

\paragraph{Massless particles  in $5D$:} According to our discussion in section $3.2$, massive $4D$-particles, which travel on timelike paths ($ds^2 > 0$), can also move on null paths in $5D$ provided the extra dimension is spacelike ($\epsilon = -1$). The bulk geodesic motion of a massless  particle ($M_{(5)} = 0$) with $dy/ds \neq 0$ is observed in $4D$ as the motion of a massive particle ($m_{0} \neq 0$) under the influence of the force given by (\ref{new def for force per unit mass contr}) and/or (\ref{new def for force per unit mass cov}) with 
\begin{equation}
\label{P4 cov for null geod}
\frac{dy}{ds} = \pm \frac{\sqrt{\Omega}}{\Phi}.
\end{equation}
\subsection{$\Phi = const$}
In brane-world theory and STM, many authors choose  to work in a Gaussian normal coordinate system based on our brane/spacetime.  This choice might be convenient because it makes $\Phi = 1$, but is not necessary\footnote{The  choice $\Phi = 1$ is not a requirement of the field equations, it is an external condition, namely, $A^{B} = {\psi}_{;C}^{B}{\psi}^{C} = 0$. In brane theory  a variable scalar field $\Phi$ entails the  possibility  of variable fundamental physical ``constants" \cite{NewVar}.}. We note that in this case  the quantities $F^{\mu}/{m_{0}}$ and $Du^{\mu}/ds$ yield the same result. This coincidence, however,  does {\em not} mean that $Du^{\mu}/ds$ represents the correct definition of the extra force when $\Phi = const$. This is  because even now $Du^{\mu}$ does not behave like a ``regular" four-vector. Indeed, {\em any} four-vector $A^{\mu}$ {\em must} satisfy the relation \cite{Landau and Lifshitz}
\begin{equation}
\label{rel. from Landau}
D A_{\mu} = g_{\mu\nu}D A^{\mu}.
\end{equation}
 On the other hand, $Du_{\mu} \neq g_{\mu\nu}Du^{\nu}$,  {\em regardless} of the choice of $\Phi$. 
 
\section{The extra force in the Brane-world scenario}
In order to evaluate the observed quantities in $4D$ we have to identify the metric of the spacetime. However, there are distinct approaches to  determine the $4D$-geometry from a given five-dimensional manifold. In this section we examine the mass and extra force in spacetime as prescribed by the brane-world scenario.  Our  purpose is to show that, in brane-world models with ${\bf Z}_{2}$-symmetry,
the extra force (\ref{new def for force per unit mass contr}) is continuous and well defined across the brane. 
We  illustrate this result with an example.  

In the brane-world scenario our spacetime is identified with a 
singular hypersurface (or $3$-brane), say $\Sigma$, orthogonal  to the $5D$ vector field 
${\psi^A}= (0, 0, 0, 0, {\Phi}^{- 1})$. 
The effective equations for gravity on a $3$-brane were obtained by Shiromizu {\it et al} \cite{Shiromizu}. In their approach the  physical metric $g_{\mu\nu}$ is identified with the induced metric $\gamma_{\mu\nu}$ (this is equivalent  to setting $\Omega = 1$) on the brane, which  is {\em fixed} at some $y = y_{0}$. 

The extra force (\ref{new def for force per unit mass contr}) has a term which is proportional to the first derivatives of the metric with respect to the extra coordinate. These derivatives can be written 
in terms of $K_{\alpha\beta}$, the extrinsic curvature of hypersurfaces $y = const$. Namely, 
\begin{equation}
\label{extrinsic curvature}
K_{\alpha\beta} = \frac{1}{2}{\cal{L}}_{\psi}g_{\alpha\beta} = 
\frac{1}{2\Phi}\frac{\partial{g_{\alpha\beta}}}{\partial y},\;\;\; K_{A4} = 0.
\end{equation}
In the brane-world scenario the 
metric is continuous across $\Sigma$, but the extrinsic curvature $K_{\mu\nu}$ 
is discontinuous. In view of this, the general belief  is that in this scenario (\ref{new def for force per unit mass contr}) cannot be implemented directly. Some authors argue that the effective equations in $4D$ should be obtained by taking the  mean values of the extrinsic curvature across $\Sigma$. 

However, for the calculation of the force the important term   is the product of the extrinsic curvature times $dy/ds$, not $K_{\mu\nu}$ alone, i.e.\footnote{As mentioned above, in this approach $g_{\mu\nu} = \gamma_{\mu\nu}$. Therefore, in this section we set $\Omega = 1$.}, 
\begin{equation}
\label{force in a brane universe}
\frac{1}{m_{0}}F^{\mu} = \epsilon \Phi \Phi^{\mu}\left(\frac{dy}{ds}\right)^2 + \Phi u^{\lambda}\left(u^{\mu}u^{\rho} - 2 g^{\mu\rho}\right)K_{\rho \lambda}\left(\frac{dy}{ds}\right).
\end{equation}
Most brane-world models assume that the universe is invariant under the ${\bf Z}_{2}$ transformation $y \rightarrow - y$, 
about our brane \cite{Maartens2}-\cite{Deruelle and Katz}.  Namely, 
\begin{eqnarray}
\label{Z2-symmetric metric}
d{\cal S}^2 &=& g_{\mu\nu}(x^{\rho}, + y)dx^{\mu}dx^{\nu} + 
\epsilon \Phi^2(x^{\rho}, + y) dy^2, \;\;\; for\;\; y \geq 0 \nonumber \\
d{\cal S}^2 &=& g_{\mu\nu}(x^{\rho}, - y)dx^{\mu}dx^{\nu} + 
\epsilon \Phi^2(x^{\rho}, - y) dy^2, \;\;\; for\;\;  y \leq 0.
\end{eqnarray}
Thus
\begin{equation}
\label{Z2 symmetry} 
K_{\mu\nu}\mid_{{\Sigma}^{+}} = - K_{\mu\nu}\mid_{{\Sigma}^{-}}.
\end{equation}
Let us now consider the bulk geodesic motion of test particles. It  can be studied by means of the Hamilton-Jacobi equation, which in $5D$ is given by
\begin{equation}
\label{HJ equation}
\gamma^{AB}\left(\frac{\partial S}{\partial x^A}\right)\left(\frac{\partial S}{\partial x^B}\right) = M_{(5)}^2,
\end{equation}
where $S$ is the five-dimensional action and the metric is given by (\ref{Z2-symmetric metric}). It is clear that the solution  of this equation in the bulk satisfies\footnote{Since the action depends only on the coordinates of the particle in $5D$, it follows that $P_{A} = - \partial S/\partial x^A$ is {\em independent} of the parameterization along the bulk geodesic.}
\begin{eqnarray}
\label{HJ in the bulk}
S^{(+)} &=& S(x^{\mu}, + y), \;\;\;\; for \;\; y \geq 0 \nonumber \\
S^{(-)} &=& S(x^{\mu}, - y), \;\;\;\; for \;\; y \leq 0.
\end{eqnarray}
The covariant components of the four-momentum, according to (\ref{identification of p with P cov}), are given by 
\begin{equation}
p_{\mu} = P_{\mu} = - \left(\frac{\partial S}{\partial x^{\mu}}\right)_{|\Sigma}.
\end{equation}
They do not depend  on whether we use $S^{(+)}$ or $S^{(-)}$ for their calculation. However, $P_{4}$ does depend on that; it changes its sign across the brane. Namely, since 
\begin{equation}
P^{(\pm)}_{4} = - \frac{\partial S^{(\pm)}}{\partial y},
\end{equation}
from (\ref{HJ in the bulk}) it follows that $P^{(+)}_{4} = - P_{4}^{(-)}$. Now, using  $P^{A} = \gamma^{AB}P_{B}$ and $P^{A} = M_{(5)} dx^A/d{\cal{S}}$, or $P^{A} = dx^A/d{\lambda}$ for $M_{(5)} = 0$, we get
\begin{equation}
\frac{dy}{ds} = \left(\frac{\gamma^{44}u^{0}}{\gamma^{00}P_{0}}\right)P_{4}.
\end{equation}
Since the metric is continuous across the brane, and the four-momentum as well as the  four-velocity  are independent on which side of the brane we are using, it follows that 
\begin{equation}
\left(\frac{dy}{ds}\right)_{|{\Sigma^{+}}} = - \left(\frac{dy}{ds}\right)_{|{\Sigma^{-}}}.
\end{equation} 
Therefore, in a ${\bf Z}_{2}$-symmetric universe the product  $K_{\mu\nu}dy/ds$  is continuous across the brane, viz,  
\begin{equation}
K_{\mu\nu}(\Sigma^+)\left(\frac{dy}{ds}\right)_{|{\Sigma^{+}}} = K_{\mu\nu}(\Sigma^-)\left(\frac{dy}{ds}\right)_{|{\Sigma^{-}}}.
\end{equation}
This means that the force (\ref{new def for force per unit mass contr}) and/or (\ref{force in a brane universe}
) is perfectly well defined in a ${\bf Z}_{2}$-symmetric universe, i.e., we get the same result regardless of  whether we calculate it from ``above" or ``bellow" the brane.

\subsection{Homogeneous cosmology in brane-world}

We now study the geodesic motion in a five-dimensional bulk space with three-dimensional isotropy and homogeneity. Our goal here is to provide an explicit example of the above discussion. The metric  may be written as 
\begin{equation}
\label{metric for a flat Friedmann universe}
d{\cal{S}}^2 = N^2(t, y)dt^2 - A^2(t, y)\left[dr^2  + r^2\left(d\theta^2 + \sin^2\theta d\phi^2\right)\right] + \epsilon \Phi^2(t, y) dy^2,
\end{equation}
 where $y$ is the coordinate along the  extra-dimension and $t, r, \theta$ and $\phi$ are the usual coordinates for a spacetime with spherically symmetric spatial sections. In spherically symmetric fields test particles move on  a single ``plane" passing through the center. We take this plane  as the $\theta = \pi/2$ plane. Thus, the Hamilton-Jacobi equation (\ref{HJ equation}) for the metric (\ref{metric for a flat Friedmann universe}) is 
\begin{equation}
\label{HJ}
\frac{1}{N^2}\left(\frac{\partial S}{\partial t}\right)^2 -  \frac{1}{A^2}\left[\left(\frac{\partial S}{\partial r}\right)^2 + \frac{1}{r^2}\left(\frac{\partial S}{\partial \phi}\right)^2\right] + \frac{\epsilon}{\Phi^2}\left(\frac{\partial S}{\partial y}\right)^2 = M_{(5)}^2.
\end{equation}
Since $\phi$ is a cyclic coordinate, it is clear that the action separates as
\begin{equation}
S = S_{1}(t, y) + S_{r}(r) + L \phi,
\end{equation}
where $L$ is the angular momentum.  
Thus, we obtain
\begin{equation}
\label{S1}
\frac{1}{N^2}\left(\frac{\partial S_{1}}{\partial t}\right)^2  - \frac{k^2}{A^2} + \frac{\epsilon}{\Phi^2}\left(\frac{\partial S_{1}}{\partial y}\right)^2 = M_{(5)}^2,
\end{equation}
and
\begin{equation}
\left(\frac{dS_{r}}{dr}\right)^2 + \frac{L^2}{r^2} = k^2 \geq 0,
\end{equation}
where $k$ is the separation constant. 

If $k = 0$, then the particle in its five-dimensional motion remains at rest in space. In this case $u^{\mu} = \delta^{\mu}_{0}/N$, and 
\begin{equation}
F^{\mu}_{\Phi \perp} = F^{\mu}_{g \perp} = 0.
\end{equation}
Consequently, in this situation only the extra force $F^{\mu}_{\parallel} = u^{\mu}dm_{0}/ds$ would be observable in $4D$. In general, in any other case with $k \neq 0$ the forces  $F^{\mu}_{\Phi \perp}$, $F^{\mu}_{g \perp}$ will be non-zero.  

In order to illustrate 
 the equations for mass and force, we  consider  the  Ricci-flat five-dimensional metric 
\begin{equation}
\label{Ricci flat de Sitter metric}
d{\cal{S}}^2 = \frac{\Lambda y^2}{3}\left\{dt^2 - e^{2\sqrt{\Lambda/3}t}\left[dr^2 +r^2\left(d\theta^2 + \sin^2\theta d\phi^2\right)\right]\right\} - dy^2. 
\end{equation}
This metric belongs to the family of separable solutions found by the present author \cite{JPdeL 1}. It  exhibits the distinctive features in  cosmology  and  is simple enough as to allow the integration of (\ref{S1}).

Here we cannot set the brane at $y = 0$. We set it at $y = y_{0} = \sqrt{3/\Lambda}$ and impose the ${\bf Z}_{2}$ symmetry under the transformation $y \rightarrow y_{0}^2/y$ (see \cite{Youm2} and references therein). The appropriate bulk background is 
\begin{equation}
\label{Ricci flat de Sitter metric above}
d{\cal{S}}^2 = \frac{y^2}{y_{0}^2}\left\{dt^2 - e^{2\sqrt{\Lambda/3}t}\left[dr^2 +r^2\left(d\theta^2 + \sin^2\theta d\phi^2\right)\right]\right\} - dy^2,  
\end{equation}
and 
\begin{equation}
\label{Ricci flat de Sitter metric bellow }
d{\cal{S}}^2 = \frac{y^{2}_{0}}{y^2}\left\{dt^2 - e^{2\sqrt{\Lambda/3}t}\left[dr^2 +r^2\left(d\theta^2 + \sin^2\theta d\phi^2\right)\right]\right\} - \frac{y_{0}^4}{y^{4}}dy^2,
\end{equation}
for $y \geq 0$ and $y \leq 0$, respectively. The metric at the brane, located at $y = y_{0}$, is the usual de Sitter metric in $4D$, 
 \begin{equation}
\label{de Sitter metric}
ds^2 = dt^2 - e^{2\sqrt{\Lambda/3}t}\left[dr^2 +r^2\left(d\theta^2 + \sin^2\theta d\phi^2\right)\right].  
\end{equation}
Thus, 
\begin{equation}
\label{K+ and K- for non-null geod in the de Sitter metric}
K_{tt}^{+} = \frac{1}{y_{0}}, \;\; K_{tt}^{-} =  - \frac{1}{y_{0}}.
\end{equation}
\subsubsection{Non-null bulk geodesics}
Let us consider the five-dimensional motion with $k = 0$. In this way we  isolate the effects of the extra dimension from the effects due to the motion in spacetime.  From (\ref{S1}) (with $\epsilon = -1$), using (\ref{Ricci flat de Sitter metric above}) and (\ref{Ricci flat de Sitter metric bellow }) we get\footnote{The sign in $S$ is chosen in such a way that the energy be positive, viz., $P_{0} = - \partial S/\partial t > 0$.}
\begin{equation}
\label{sol of S1+}
S_{1}^{(+)} = - M_{(5)} y \sinh\sqrt{\frac{\Lambda}{3}}t,\;\;\;S_{1}^{(-)} = -  M_{(5)} \frac{y_{0}^2}{y} \sinh\sqrt{\frac{\Lambda}{3}}t.
\end{equation}
The four-momentum is a well defined quantity, viz.,
\begin{equation}
\label{four-momentum for brane sol. with non-null geod}
p_{\mu} = - \left(\frac{\partial S}{\partial x^{\mu}}\right)_{|\Sigma} = \delta_{\mu}^{0}M_{(5)}\cosh\sqrt{\frac{\Lambda}{3}}t.
\end{equation}
 Now using 
\begin{equation}
\label{5D contr momentum in terms of 5D metric and action}
P^A = M_{(5)}\frac{dx^A}{d{\cal{S}}} = \gamma^{AB}P_{B} = - \gamma^{AB}\frac{\partial S}{\partial x^B},
\end{equation}
we obtain
\begin{equation}
\label{the momentum}
P^{A} = \left(M_{(5)}\frac{y_{0}}{y}\cosh\sqrt{\frac{\Lambda}{3}}t,\;\;\; 0,\;\;\; 0,\;\;\; 0,\;\;\; - M_{(5)}\sinh \sqrt{\frac{\Lambda}{3}}t\right),
\end{equation}
for $y \geq 0$, and 
\begin{equation}
P^{A} = \left(M_{(5)}\frac{y}{y_{0}}\cosh\sqrt{\frac{\Lambda}{3}}t,\;\;\; 0,\;\;\; 0,\;\;\; 0,\;\;\; + M_{(5)}\frac{y^2}{y_{0}^2}\sinh \sqrt{\frac{\Lambda}{3}}t\right),
\end{equation}
for $y \leq 0$.
From these expressions we get
\begin{equation}
\label{dy/ds in brane theory}
\left(\frac{dy}{dt}\right)_{|\Sigma^{+}} =  - \tanh\sqrt{\frac{\Lambda}{3}}t, \;\;\;\left(\frac{dy}{dt}\right)_{|\Sigma^{-}} =  + \tanh\sqrt{\frac{\Lambda}{3}}t.
\end{equation}
In the present case $u^{\mu} = \delta^{\mu}_{0}$, so that $dt/ds = 1$. Thus, from (\ref{K+ and K- for non-null geod in the de Sitter metric})
we obtain
\begin{equation}
K_{tt}(\Sigma^+)\left(\frac{dy}{ds}\right)_{|{\Sigma^{+}}} = K_{tt}(\Sigma^-)\left(\frac{dy}{ds}\right)_{|{\Sigma^{-}}} = - \sqrt{\frac{\Lambda}{3}}\tanh\sqrt{\frac{\Lambda}{3}}t.
\end{equation}
Consequently, the extra force as observed on the brane (\ref{force in a brane universe}) is given by 
\begin{equation}
\label{extra force, brane theory, non-null geod}
\frac{F^{\mu}}{m_{0}} = \delta^{\mu}_{0}\sqrt{\frac{\Lambda}{3}}\tanh\sqrt{\frac{\Lambda}{3}}t.
\end{equation}
The rest mass measured on the brane is given by (\ref{relation between the rest mass in 4D and 5D}) as 
\begin{equation}
\label{rest mass, brane theory, non-null geod}
m_{0} = M_{(5)}\cosh\sqrt{\frac{\Lambda}{3}}t.
\end{equation}
It is clear that these expressions are consistent with (\ref{variation of the effective mass}).

\subsubsection{Null bulk geodesics}
If $M_{(5)} = 0$, then the bulk motion is along null geodesics. In this case the $5D$ action is given by
\begin{equation}
\label{null action}
S_{1(Null)}^{(+)} = C y e^{ -  \sqrt{\Lambda/3}t},\;\;\;S_{1(Null)}^{(-)} = C\frac{y_{0}^2}{ y} e^{ -  \sqrt{\Lambda/3}t}.
\end{equation}
 where $C$ is a constant of integration. The corresponding four-momentum as observed on the brane (located at $y = y_{0} = \sqrt{3/\Lambda}$) is
\begin{equation}
p_{\mu} = \delta_{\mu}^{0}C e^{ -  \sqrt{\Lambda/3}t}.
\end{equation} 
Consequently, for the mass we obtain
\begin{equation}
\label{mass for null geod}
m_{0} = p_{\mu}u^{\mu} = C e^{ -  \sqrt{\Lambda/3}t}.
\end{equation}
The extra force, per unit mass, due to variation of rest mass is 
\begin{equation}
\label{force for null geod seen from the brane}
\frac{F^{\mu}}{m_{0}} = - \delta^{\mu}_{0}\sqrt{\frac{\Lambda}{3}}.
\end{equation}
Identical results can be obtained from an analysis similar to the one in section $5.1.1$. Here $(dy/ds)_{|\Sigma^{+}} = - (dy/ds)_{|\Sigma^{-}} = + 1$. Thus, using (\ref{K+ and K- for non-null geod in the de Sitter metric}), from (\ref{force in a brane universe}) we recover (\ref{force for null geod seen from the brane}). We also note that the mass of the particle can be obtained by evaluating  (\ref{rel between m, P4 and M}), with $M_{(5)} = 0$, from either side of the brane.

\section{The extra force in STM}

In this section we compare the rest mass and extra force as predicted  by STM and  brane theory. We will show that, although these theories  give distinct prescriptions for the geometry of the spacetime,  they lead to identical results for the mass and force as observed in $4D$. 

In STM our four-dimensional world is embedded in a five-dimensional spacetime, which is a solution of the five-dimensional Einstein's equations in vacuum.  The extra dimension is not assumed to be compactified, which allows us to obtain the properties of matter as a consequence of the large extra dimension.

An important  similarity between brane theory and STM is that in both schemes our four-dimensional spacetime is identified with a {\em fixed} hypersurface $\Sigma$ (defined by the equation $y = y_{0} = constant$), and the  metric in $4D$ is taken to be the induced one, viz., $g_{\alpha\beta}(x^{\mu}) = \gamma_{\alpha\beta}(x^{\mu}, y _{0})$. 
The main difference is that this hypersurface is singular in brane theory and non-singular in STM. However,  the effective matter content of  spacetime is the same whether calculated from STM equations or from the ${\bf Z}_{2}$-symmetric brane perspective.

Although STM and brane theory have different physical motivations for the introduction of a large extra dimension, they 
share the same working scenario and are equivalent in many respects. In particular, STM  includes the so-called local high-energy corrections, and non-local Weyl corrections typical of brane-world scenarios \cite{STMBrane}.

We now proceed to show that both prescriptions, brane-theory and STM, lead to the same expressions for the rest mass and extra force as observed in $4D$. 
This is clear from the fact that in both theories all relevant quantities for the calculation of the extra force  (\ref{new def for force per unit mass contr}) are continuous across $\Sigma$. Indeed, in STM there are no discontinuities, and 
in brane theory $K_{\mu\nu}dy/ds$ is continuous across the brane, despite of the fact that  each quantity; $K_{\mu\nu}$ and $dy/ds$ is discontinuous separately. 

As an illustration, let us again consider the $5D$ metric  (\ref{Ricci flat de Sitter metric}). In STM the bulk metric is the same in both sides of $\Sigma$, which we locate at  $y = y_{0} = \sqrt{3/\Lambda}$. Thus, in our calculation  we can use either (\ref{Ricci flat de Sitter metric above}) or (\ref{Ricci flat de Sitter metric bellow }). If we choose (\ref{Ricci flat de Sitter metric above}), then $(\partial g_{tt}/\partial y)_{\Sigma} = 2\sqrt{\Lambda/3}$ (in this prescription $\Omega = 1$). The results observed in $4D$ depend on whether the motion in the bulk is along non-null or null geodesics. 

In the case of non-null bulk geodesic motion we have $(dy/ds)_{\Sigma} = - \tanh{\sqrt{\Lambda/3}t}$. Thus, when evaluating the mass from (\ref{relation between the rest mass in 4D and 5D}) and the extra force from (\ref{new def for force per unit mass contr}),  we obtain the same results as in brane theory, namely (\ref{extra force, brane theory, non-null geod}) and (\ref{rest mass, brane theory, non-null geod}). In the case of null bulk geodesics we have $(dy/ds)_{\Sigma} = 1$. Therefore, we recover the results (\ref{mass for null geod}) and (\ref{force for null geod seen from the brane}) obtained on the brane. 

Thus, the mass and extra force perceived by an observer in $4D$ are independent of whether the bulk geodesic motion is interpreted on the non-singular hypersurface $\Sigma$ of STM or on the singular hypersurface (located at $y = y_{0}$) of brane theory. These two $5D$ theories produce indistinguishable results for test particles as observed in $4D$.

\section{The extra force in other non-compact theories}

The aim of this section is to show, by means of an explicit example, how the results for mass and force as observed in $4D$  severely depend on the way we separate the spacetime from the extra dimension. 

With this aim, here we consider two different approaches, which are alternative interpretations  of STM. They hold a different view regarding the identification of spacetime. Namely, in these approaches the geometry of the $4D$ spacetime is identified with the {\em entire} foliation orthogonal  to the $5D$ vector field 
${\psi^A}= (0, 0, 0, 0, {\Phi}^{- 1})$, instead of a fixed hypersurface $\Sigma$.

In order to illustrate these interpretations, we go back to the $5D$ Ricci-flat manifold (\ref{Ricci flat de Sitter metric}). First, we revisit the bulk geodesic motion with $k = 0$. Second, we interpret  the bulk geodesic motion as observed in $4D$.

Since there are no discontinuities, the action throughout the bulk is given by 
\begin{equation}
S_{1} = - M_{(5)}y \sinh\sqrt{\frac{\Lambda}{3}t}.
\end{equation}
From (\ref{5D contr momentum in terms of 5D metric and action}) it follows that
\begin{equation}
\label{five velocity}
U^A = \left(\frac{dt}{d{\cal{S}}}, 0,0,0, \frac{dy}{d{\cal{S}}}\right) = \left(\sqrt{\frac{3}{\Lambda}}\frac{1}{y}\cosh\sqrt{\frac{\Lambda}{3}}t, \;\;\; 0,\;\;\;0,\;\;\;0,\;\;\; - \sinh\sqrt{\frac{\Lambda}{3}}t\right).
\end{equation}
Consequently,
\begin{equation}
\label{equation for y in the cosmological example}
y = \frac{{\bar{y}}_{0}}{\cosh\sqrt{\frac{\Lambda}{3}}t},
\end{equation}
where ${\bar y}_{0}$ is a constant of integration. Therefore, according to (\ref{identification of p with P cov}) the four-momentum observed in this approach is given by 
\begin{equation}
\label{p no-null}
p_{\mu} = ({\bar{y}}_{0}M_{(5)}\sqrt{\frac{\Lambda}{3}},0, 0, 0).
\end{equation}
In the case of null geodesics in $5D$ the action is given by 
\begin{equation}
\label{null action}
S^{(Null)}_{1} = C y e^{ - \alpha \sqrt{\Lambda/3}t},
\end{equation}
 where $\alpha = \pm 1$ and $C$ is a constant of integration. In (\ref{5D contr momentum in terms of 5D metric and action}) we replace $M_{(5)}d/d\cal{S}$ by $d/d\lambda$, where $\lambda $ is the parameter along the null geodesic, and obtain 
\begin{equation}
P^{A} = \left(\frac{dt}{d \lambda}, 0, 0, 0, \frac{dy}{d \lambda}\right) = \left(\sqrt{\frac{3}{\Lambda}}\frac{\alpha C}{y}e^{ - \alpha \sqrt{\Lambda/3}t},\;\;\; 0,\;\;\; 0,\;\;\; 0,\;\;\; C e^{ - \alpha \sqrt{\Lambda/3}t}\right).
\end{equation}
In this case
\begin{equation}
\label{y Null}
y^{(Null)} = {\bar{y}}_{0}e^{\alpha \sqrt{\Lambda/3}t}, 
\end{equation}
and  the four-momentum as observed in $4D$ is
\begin{equation}
\label{p Null}
p_{\mu} = ( \alpha C {\bar{y}}_{0}\sqrt{\frac{\Lambda}{3}}, 0, 0, 0).
\end{equation}
It is important to note that  (\ref{y Null}) and (\ref{p Null}) are {\em independent} of the choice of geodesic parameter $\lambda$.

For the interpretation of the bulk geodesic motion as observed in $4D$ we have to identify the metric of the physical spacetime. We will consider two approaches.

 \subsection{Canonical approach}
In this approach the metric in the bulk is simplified by using all five available coordinate degrees of freedom to set $\gamma_{\mu 4} = 0$ and $\Phi = 1$. Besides, the  physical metric in $4D$ is assumed to be conformally related to the induced one. The warp factor is taken as $\Omega = (y/L)^2$. Namely, 
\begin{equation}
\label{canonical metric}
d{\cal{S}}^2 = \frac{y^2}{L^2}g_{\mu\nu}(x^{\alpha}, y)dx^{\mu}dx^{\nu} - dy^2.
\end{equation} 
 This metric is usually called canonical metric \cite{Wesson book}. Here $L$ is a constant of length, which in cosmological solutions is identified with the cosmological constant via $L = \sqrt{3/\Lambda}$.

We note that 
the Ricci-flat metric (\ref{Ricci flat de Sitter metric})  has the canonical form (\ref{canonical metric}). Thus, in the canonical metric approach the geometry of the spacetime is determined by (\ref{de Sitter metric}) and 
the warp factor is $\Omega = \Lambda y^2/3$. Since $\partial g_{\mu\nu}/\partial y = 0$, from 
(\ref{new def for force per unit mass contr}) it follows that 
\begin{equation}
F^{\mu} = 0,\;\;\;(\mu = 0,1,2,3),
\end{equation}
in this interpretation.  This is consistent with the fact that here the rest mass is constant, which is a consequence of (\ref{variation of the effective mass}). 
In order to get $m_{0}$ we can use (\ref{p no-null}). Namely, 
\begin{equation}
\label{mass first case}
m_{0} = p_{\mu}u^{\mu} = {\bar{y}}_{0} M_{(5)}\sqrt{\frac{\Lambda}{3}}. 
\end{equation}
This also can be obtained by direct substitution of (\ref{equation for y in the cosmological example}) into (\ref{relation between the rest mass in 4D and 5D}).

 If $M_{(5)} = 0$, then the  motion is along null geodesics in $5D$.  From (\ref{P4 cov for null geod}) and (\ref{y Null}) we obtain $dt/ds = \alpha$. Therefore, from (\ref{p Null})
\begin{equation}
  m_{0} = {\bar{y}}_{0} C \sqrt{\frac{\Lambda}{3}},
\end{equation}
\subsection{Induced-metric approach}
In this approach the metric of the spacetime is identified with the one induced on the set of  hypersurfaces orthogonal to the $5D$ vector field
${\psi^A}$, which  is given by 
\begin{equation}
g_{\mu\nu}(x^{\rho}, y) = h_{\mu}^{A}h_{\nu}^{B}\gamma_{AB}(x^{\rho}, y),
\end{equation}
where $h_{AB}$ is the projector introduced in (\ref{projector}). We note that in brane theory as well as STM the spacetime is fixed at some $y = y_{0} = const$. In the present approach, which is also called ``foliating" approach \cite{Romero2}, the geometry of the spacetime is determined by the whole family of orthogonal hypersurfaces. 

In the case under consideration  
the metric of the hypersurfaces orthogonal to the $5D$ vector field
${\psi^A}= (0, 0, 0, 0, 1)$ is given by  
\begin{equation}
\label{metric in 4D in the foliating approach}
ds^2 = \frac{\Lambda y^2}{3}\left\{dt^2 - e^{2\sqrt{\Lambda/3}t}\left[dr^2 +r^2\left(d\theta^2 + \sin^2\theta d\phi^2\right)\right]\right\}. 
\end{equation}
For this metric $u_{\mu} = \delta_{\mu}^{0}\sqrt{\Lambda/3}y$. Since $p_{\mu} = m_{0}u_{\mu}$, from (\ref{p no-null}) it follows that
\begin{equation}
\label{mass in the foliating approach}
m_{0} =  M_{(5)} \cosh \sqrt{\frac{\Lambda}{3}}t,
\end{equation}
where we have evaluated $y$ along the trajectory by using (\ref{equation for y in the cosmological example}). In this approach  $\Omega = 1$, $dt/ds = (\sqrt{3/\Lambda}/{\bar{y}}_{0})\cosh{\sqrt{\Lambda/3}t}$,  and $dy/ds = - \tanh\sqrt{\Lambda/3}t$. Consequently, from our general equation (\ref{relation between the rest mass in 4D and 5D}) we get the same result as in (\ref{mass in the foliating approach}), as expected.

Taking derivatives in (\ref{mass in the foliating approach}) we get
\begin{equation}
\label{extra 4D force for non-null 5D geod motion in the foliating approach}
\frac{F^{\mu}_{\parallel}}{m_{0}} = \frac{\delta^{\mu}_{0}}{2 {\bar{y}}_{0}^2}\sqrt{\frac{3}{\Lambda}} \sinh 2 \sqrt{\frac{3}{\Lambda}}t.
\end{equation}
It is easy to verify that this  result is  consistent with (\ref{extra force parallel to four velocity, new definition}). 

\medskip

In order to avoid misunderstanding,  we should mention  that the $4D$ metric contains no $y$. It is obtained from (\ref{metric in 4D in the foliating approach}) after we substitute (\ref{equation for y in the cosmological example})  into it. Namely,   
\begin{equation}
\label{metric in 4D in the foliating approach from non-null geod}
ds^2 = \frac{\Lambda {\bar{y}}_{0}^{2}}{3 \cosh^2{\sqrt{\Lambda/3}t}}\left\{dt^2 - e^{2\sqrt{\Lambda/3}t}\left[dr^2 +r^2\left(d\theta^2 + \sin^2\theta d\phi^2\right)\right]\right\}. 
\end{equation}
This line element is distinct from the one  in the brane-world and STM interpretation, which is given by (\ref{de Sitter metric}). As a consequence, (\ref{rest mass, brane theory, non-null geod}) and (\ref{mass in the foliating approach}) are {\em distinct} functions of the proper time. 

\medskip

For the case of null geodesic motion in the bulk metric (\ref{Ricci flat de Sitter metric}), in the induced-metric approach we have $dy^2 = ds^2$. Taking  $dy/ds = 1$, from (\ref{y Null}) we get $dt/ds = (\sqrt{3/\Lambda}/\alpha {\bar{y}}_{0})e^{- \alpha\sqrt{\Lambda/3}t}$. Thus,  using  (\ref{p Null}) we obtain
\begin{equation}
\label{4D mass for null 5D geod in the foliating approach}  
m_{0}  = C e^{- \alpha\sqrt{\Lambda/3}t} = \frac{(\alpha C {\bar{y}}_{0})}{s},
\end{equation}
where $s$ is the proper time. Taking derivatives we find the extra force as
\begin{equation}
\label{extra 4D force for null 5D geod motion in the foliating approach}
\frac{F^{\mu}_{\parallel}}{m_{0}} =  -  \delta^{\mu}_{0}\sqrt{\frac{3}{\Lambda}}\frac{1}{s^2}.
\end{equation}
This can be easily corroborated from (\ref{extra force parallel to four velocity, new definition}). The geometry in $4D$, is given by (\ref{metric in 4D in the foliating approach}) evaluated at $y = y^{(Null)}$  from (\ref{y Null}). 

The discussion of this section clearly shows that the rest mass as well as the  extra force as observed in $4D$ depend on (i) the method  we use to identify  the $4D$ metric from the $5D$ one, (ii) the nature of the geodesic motion in $5D$ and (iii) the motion in $3$-space. 

\section{Summary and final comments}

The aim of this work has been to present a clear and general discussion of how an observer in $4D$ interprets the geodesic motion in a five-dimensional bulk space. Here we have provided a unified methodology
for the discussion of the mass and extra force as observed in $4D$. Our method presents a number of advantages over other studies in the literature. First, it can successfully be applied  to compactified Kaluza-Klein theory, brane world, STM,  and other non-compact  theories in $5D$. Second,  the whole discussion is free of the subtle details associated with the choice of affine parameters used to describe the motion in $4D$ and $5D$. Third, it works equally well for non-null and null geodesics in the bulk (the latter involves the  change of $dy/ds$ by $\pm \sqrt{\Omega}/\Phi$. 

In the scenario of compactified Kaluza-Klein theory (with the cylinder condition) the extra force reduces to 
\begin{equation}
\label{new def for force per unit mass cov for KK}
\frac{1}{m_{0}}F_{\mu} =  \frac{\epsilon \Phi}{\Omega} \Phi_{\mu}\left(\frac{dy}{ds}\right)^2, \;\;\; \frac{1}{m_{0}}F_{\mu} = - \frac{\Phi_{\mu}}{\Phi},
\end{equation}
for non-null and null bulk geodesics, respectively. We remind the reader that in the case of null bulk geodesics the extra coordinate has to be spacelike ($\epsilon = - 1$), otherwise the particles observed in $4D$ are massless. The above equations show that the existence of an extra force is not a prerogative of theories with large extra dimensions like brane theory and STM. Conversely, there are $5D$ metrics with explicit dependence on the extra coordinate, which show constant rest mass and  no extra force, when they are interpreted  in the context of the brane/STM scenario. In these metrics the constancy of the rest mass is a consequence  of the mutual cancelation of the mass-change induced by the term $(\partial g_{\mu\nu}/\partial y)u^{\mu}u^{\nu}$ and the one induced by the scalar field. 

In the brane world scenario with ${\bf Z}_{2}$-symmetry, we have shown that the extra force  
is continuous and well defined across the brane. 
This is an effect of the required symmetry. In fact, in such a scenario the momentum component along the extra dimension changes its sign  across the brane, 
which effectively compensates the discontinuity of the extrinsic curvature. This is an important result because if our universe is described by the brane world scenario, then it has to have ${\bf Z}_2$ symmetry. Indeed, if the ${\bf Z}_2$-symmetry is dropped, then there is an extra term in the Friedmann equation \cite{Vernon}. This  term is constrained by the condition that standard cosmology is in place by the time of nucleosynthesis.
In other words,  the effects associated with the lack of ${\bf Z}_2$ symmetry must decrease with time. Which means that the extra term  should be small enough at the time of nucleosynthesis and negligible today. This is why 
brane-world models {\em without} ${\bf Z}_2$ symmetry (at late times) seem to be of no observational significance today. 

In the original interpretation of STM, our four-dimensional spacetime was identified with a {\em fixed} hypersurface $\Sigma$ (defined by the equation $y = y_{0} = constant$), and the  metric in $4D$ was taken to be the induced one. With this identification, we find that brane world theory as well as  STM lead to equal results for the mass and extra force as observed on the three-brane/spacetime. This means  that observations made with particles cannot help us to distinguish whether we live on a singular or regular brane/STM hypersurface. 
This result is compatible with previous investigations where we showed that these two theories are equivalent to each other, although they look very different at first sight \cite{STMBrane}. 

Subsequent interpretations of STM use the so-called canonical metric, in which the geometry of the $4D$-spacetime is taken to be conformally related  to the induced metric. This approach is {\em not} equivalent to the brane/STM scenario discussed above. This is illustrated by our example in section $7.1$, which shows that, unlike the observations made on the brane, the rest mass is constant and consequently there is no extra force. 

Another alternative approach which deserves consideration is the one where the geometry of the $4D$ spacetime is determined {\em not} by a fixed hypersurface $y = const$, but by the {\em whole} family of hypersurfaces orthogonal to the extra dimension. This brings to mind the situation where the motion of a test particle  is described from two distinct frames of reference. For instance, the comoving  and some other non-comoving frame. Certainly the observed quantities  in these frames are different. The corresponding similarity in $5D$ is clear.  The observations made in the fixed brane/STM ($y = const$) should be different from those made on the ``moving" ($y \neq const$) brane. This explains the results in  our example in section $7.2$. 

To conclude, the two leading five-dimensional theories, namely brane world and STM, predict identical results for test particles  as observed in $4D$. However, other approaches seem to be possible. The discovery of new physical phenomena,  unmistakable related to  extra dimensions, is a challenge for these theories.

\end{document}